# Hybrid High-Temperature Superconductor-Semiconductor Tunnel Diode

Alex Hayat[1,2], Parisa Zareapour[1], Shu Yang F. Zhao[1], Achint Jain[1], Igor G. Savelyev[3], Marina Blumin[3], Zhijun Xu[4], Alina Yang[4], G. D. Gu[4], Harry E. Ruda[2,3], Shuang Jia[5], R. J. Cava[5], Aephraim M. Steinberg[1,2], and Kenneth S. Burch[1]

[1]Department of Physics and Institute for Optical Sciences, University of Toronto, Toronto, Ontario M5S 1A7, Canada

[2]Centre for Quantum Information and Quantum Control University of Toronto, 60 St. George Street, Toronto ON, M5S 1A7, Canada

[3]Centre for Advanced Nanotechnology and Institute for Optical Sciences, University of Toronto, 170 College Street, Toronto M53 3E4, Canada

[4]CMP&MS Department, Brookhaven National Laboratory, Upton, New York 11973, USA

[5]Department of Chemistry, Princeton University, Princeton, New Jersey 08544, USA.

We report the demonstration of hybrid high-$T_c$-superconductor-semiconductor tunnel junctions, enabling new interdisciplinary directions in condensed matter research. The devices were fabricated by our newly-developed mechanical bonding technique, resulting in high-$T_c$-semiconductor planar junctions acting as superconducting tunnel diodes. Tunneling-spectra characterization of the hybrid junctions of $Bi_2Sr_2CaCu_2O_{8+\delta}$ combined with bulk GaAs, or a GaAs/AlGaAs quantum well, exhibits excess voltage and nonlinearity - in good agreement with theoretical predictions for a *d*-wave superconductor-normal material junction, and similar to spectra obtained in scanning tunneling microscopy. Additional junctions are demonstrated using $Bi_2Sr_2CaCu_2O_{8+\delta}$ combined with graphite or $Bi_2Te_3$. Our results pave the way for new methods in unconventional superconductivity studies, novel materials and quantum technology applications.

Superconductors enable the implementation of fast ultrasensitive detectors [1,2] and large-scale quantum computation technology [3,4]. These materials pose major scientific and technological challenges, however. A potential alternative avenue are hybrid semiconductor-superconductor devices, which have been attracting growing attention lately as they combine the controllability of semiconductor structures with the macroscopic quantum states of superconductors[5,6]. The interaction of light with semiconductor-superconductor structures has recently emerged as a new interdisciplinary field of superconducting optoelectronics, with demonstrations of light emission from hybrid light-emitting diodes [7,8] enhanced by the superconducting state [9,10], and various proposals for novel lasers [11] and quantum light sources [12,13]. These hybrid devices have also proven useful in nonlinear electronics [14,15] and infrared detection [16], taking advantage of the relatively small size of the superconducting gap in the tunneling spectrum [17]. All previously studied semiconductor-superconductor devices were based on conventional low-critical-temperature (low-$T_c$) superconductors, requiring cooling to extremely low temperatures. Moreover, the small superconducting gaps of these materials limit the energy scales over which they can be employed. High operating temperature and large $d$-wave gaps can be obtained by incorporating unconventional high-$T_c$ superconductors [18,19] that exhibit a variety of novel phenomena and provide a more practical alternative for device implementation. Furthermore, by combining high-$T_c$ materials with semiconductors, one could take advantage of mature semiconductor technology to probe the unconventional nature of high-$T_c$ superconductors [20] in hybrid tunneling junctions.

Tunneling spectroscopy is among the most widely used techniques for the study of novel materials and new phenomena in condensed matter physics [21]. Various effects have been observed with tunneling spectroscopy such as weak localization [22], superconducting gap dependence on magnetic field [23], bound states and broken symmetries in high-$T_c$ superconductors [24] as well as studies of the pseudogap, preformation of Cooper pairs [25] and electron-hole asymmetry [26]. These experiments usually require sophisticated and expensive scanning tunneling spectroscopy (STM) equipment. A simple method of constructing high-$T_c$ tunnel junctions can conceptually facilitate tunneling spectroscopy studies of these unconventional materials. Hybrid

semiconductor-high-$T_c$ optoelectronic devices such as superconducting light sources [12, 13] could help shed new light on the physics of high-$T_c$ materials by demonstrating photon-pair emission from Cooper pairs above $T_c$, which is still a highly debated issue [27]. The realization of such devices has been prevented, however, by the fact that high-$T_c$ epitaxial layers can be grown only on a very limited range of substrates.

Here we demonstrate a hybrid high-$T_c$-superconductor-semiconductor device, namely a superconducting tunnel diode constructed from $Bi_2Sr_2CaCu_2O_{8+\delta}$ (Bi-2212) combined with bulk GaAs, a GaAs/AlGaAs quantum well (QW), graphite or $Bi_2Te_3$. The devices were fabricated by our newly-developed method of mechanically-bonded planar junctions, and characterized by DC current voltage (I-V) and AC differential conductance measurements. Similar to previous work on low-$T_c$ hybrid structures, these junctions reveal nonlinear current-voltage characteristics and excess voltage [28]. This is a unique feature resulting from the gap in the superconductor quasiparticle excitation spectrum. These features as well as a broadening of the tunneling spectrum by quasiparticle desphasing are in good agreement with the *d*-wave superconductor-normal material junction theory [29]. We also show similar nonlinearity and excess voltage in graphite based junctions and junctions based on $Bi_2Te_3$ – a novel material that has recently been shown to have topologically protected surface states [30]. The fact that the agreement between theory and experiment holds in a wide array of normal materials (GaAs, graphite, $Bi_2Te_3$) with very different normal state properties, demonstrates the robustness of this technique for studying high $T_c$ superconductors.

The hybrid planar junctions were constructed by our newly developed mechanical bonding technique, enabling implementation of semiconductor/high-$T_c$ devices. This method provides a unique and simple solution for combinations of novel materials that cannot be achieved by conventional approaches. The simplicity of the fabrication method, which does not require expensive facilities, and the resulting devices, which can operate at temperatures of liquid nitrogen, make this approach technologically practical as well as greatly expand the range of possible experiments in the field of high-$T_c$ superconductivity. This was achieved using Bi-2212 crystals grown by the floating-zone method [31]. The Bi-2212 was cleaved in a nitrogen-purged dry box with adhesive tape producing atomically flat surfaces over large areas. Subsequently, it was mechanically bonded to p-doped (carrier density $> 5 \cdot 10^{18} cm^{-3}$) bulk GaAs or a GaAs/AlGaAs QW, where the doping is required to reduce the effect of the Schottky barrier. The mechanical bonding is achieved by freshly cleaving Bi-2212 in a dry atmosphere, then rapidly pressing it onto the semiconductor surface. The Bi-2212 crystals used in the measurements presented here are rectangular samples about 1mm$^2$ in area and several hundreds of μm in thickness. To confirm that surface forces are strong enough to guarantee good contact between the two materials, we first mechanically exfoliated nano-crystals of Bi-2212 [32] on top of GaAs; however, due to the difficulty of performing conductance measurements on exfoliated nano-crystals, we have chosen to focus our initial efforts on bulk crystals (Fig. 1 inset).

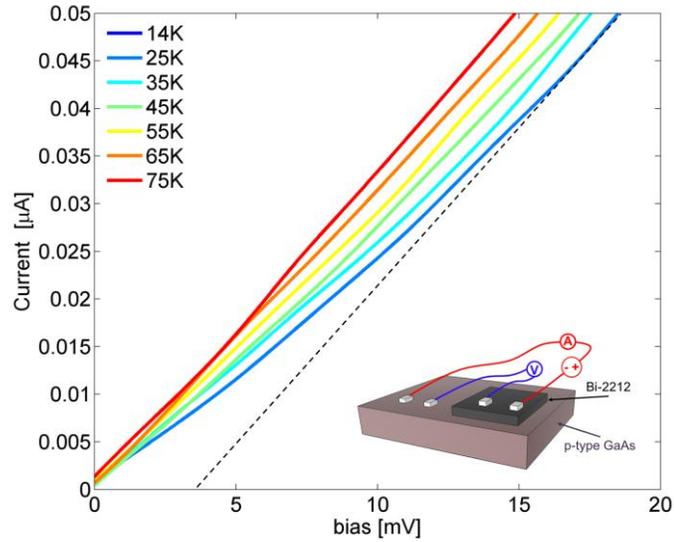

**Figure 1.** Measured DC I-V characteristics for the Bi-2212/GaAs junction at various temperatures. The dashed line indicates a linear I-V dependence coinciding with the 14K measurement at higher bias to indicate the excess voltage. The inset is a schematic drawing of the Bi-2212/GaAs device.

The larger samples should also have increased surface forces holding the GaAs and Bi-2212 together. Nonetheless, due to the increased volume to surface area ratio of the bulk crystal, we have found that it is necessary to apply GE varnish to the edges of the Bi-2212 to secure the bonding. The exact Bi-2212/GaAs contact area is difficult to determine, because the atomically flat regions are smaller than the area of the entire crystal. However, judging from our successful mechanical exfoliation of Bi-2212 on GaAs, as well as atomic force microscopy of the crystal surface, a large fraction of the Bi-2212 sample surface is in contact with the GaAs substrate.

To probe the interface with four-point differential conductance measurements, contacts were made on the top of bulk Bi-2212 and the semiconductor substrate using Cu wires and Ag epoxy. To confirm that we are only probing tunneling at the interface, different in-plane geometries of the contacts were tested, without any effect of the geometry on the measurements. All the processing steps besides crystal growth were done at room temperature, and the key requirement is to perform them in a clean and dry atmosphere – in our case in a nitrogen-purged glove box. The differential conductance spectroscopy measurements included both DC current-voltage characteristics and AC differential conductance versus voltage. The temperature-dependent measurements were performed using a liquid He flow cryostat.

The first set of measurements was performed on Bi-2212/bulk-GaAs junctions. The Bi-2212 crystals on GaAs were measured to be slightly under-doped ($T_c \sim 70K$). This can result partially from the diffusion of oxygen interstitials due to oxidation of the GaAs surface. Differential AC conductance and DC I-V characteristics at different temperatures were measured on the junction using two lock-in amplifiers (Stanford Research Systems SR810) and a DC voltage source (BK Precision 1787B). The measurements are current-polarized, where the current source is implemented using a DC voltage source with a large resistor. The AC voltage from one of the lock-in amplifiers was added to the DC output of the power supply using a shielded transformer-based adder. The AC+DC voltage was applied to the sample, and the resulting voltage drop was measured with a lock-in (the AC part) and a multimeter (the DC part). The current was converted to voltage and measured with another lock-in (AC) and a multimeter (DC) using a preamplifier (SRS 570). The small amplifier offset results in slightly non-zero current at zero bias in DC measurements, which is much smaller than the important measured values such as the current corresponding to the excess voltage.

Studying the temperature dependence of the I-V characteristics of the Bi-2212/GaAs bulk junction (Fig. 1) shows that for temperatures above $T_c$, the junction behaves almost as a normal Ohmic resistor, $R_n$. Above $T_c$, the I-V can be slightly nonlinear due the effect of the pseudogap, which is common amongst high-$T_c$ superconductors [33]. As the junction is cooled below $T_c$, the I-V characteristics become significantly nonlinear due to the reduction of the tunneling probability for single quasiparticle excitations when the superconducting gap opens up in the density of states (DOS) of Bi-2212. The nonlinear low-temperature I-V dependence becomes linear for higher bias voltages, however it is displaced from the normal I-V characteristic by the excess voltage $V_e = V - R_n I$ [28]. This displacement from normal linear I-V dependence can be extracted by extrapolating the linear part of the low-temperature curve in positive bias to the axis (Fig. 1 dashed line). The extracted excess voltage decreases with increasing temperature due to the reduction of the Bi-2212 gap.

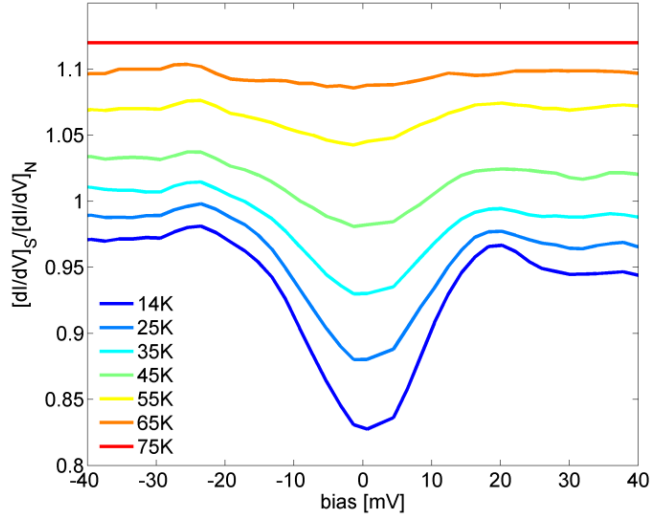

**Figure 2.** Measured AC differential conductance below $T_c$ $[dI/dV]_S$, divided by the normal state conductance $[dI/dV]_N$ for the Bi-2212/GaAs junction at various temperatures. The curves are shifted vertically for clarity by 2%, and the vertical axis corresponds to the 14K curve.

A direct measurement of the superconducting gap was performed by AC differential conductance. Below $T_c$, the ratio of the differential conductance $[dI/dV]_S$, to the normal state conductance $[dI/dV]_N$ demonstrates the temperature dependence of the superconducting gap of Bi-2212 (Fig. 2), and resembles point-contact and scanning tunneling measurements taken on bulk Bi-2212 [34,35], consistent with an underdoped sample [36]. The pair potential in Bi-2122 is *d*-wave and thus highly anisotropic in the *ab* plane [37] so in our *c*-axis tunneling experiment an averaged V-shaped gap appears with the maximal value close to 20mV (consistent with underdoped Bi-2212). The excess voltage should be proportional to, yet smaller than the maximal value observed in the V-shaped gap even at low temperatures (Fig.1, 2).

At low temperatures the depth of the gap, the sharpness of the features in the differential conductance, and the nonlinearity of the I-V characteristics are determined by the scattering rates of the quasiparticles (QP) in the normal material [38]. Carrier scattering rates in bulk GaAs are typically faster than 1 ps$^{-1}$ [39] leading to energy broadening of more than 10 meV. This significantly reduces the excess voltage in the I-V characteristics and the depth of the gap-induced dip in the differential conductance measurements.

Two-dimensional structures such as InAs/AlSb QWs have been shown to exhibit highly nonlinear I-V characteristics with low-$T_c$ Nb-based structures [40]. It has been shown recently that Dynes broadening is affected by electromagnetic dissipation [41]. We show here, that a different type of dissipation, namely inelastic scattering of QPs, has a significant effect on the tunneling spectrum. Specifically, the scattering rate is determined by the dimensionality of the semiconductor structure. QWs can have much longer carrier scattering times than those of bulk materials [42], even in the presence of inter-subband transitions [43], resulting in smaller broadening, stronger I-V nonlinearities, and larger excess voltages. In addition, by introducing a QW into the structure one gains a significant degree of freedom in tuning the overall properties of the junction, which can be crucial for device design.

In order to enhance the superconducting tunnel effect, we designed and fabricated Bi-2212/GaAs/AlGaAs junctions with a GaAs QW. The Be-doped GaAs QW (10nm thick) was grown by molecular beam epitaxy (MBE) on a Si-doped Al$_{0.25}$Ga$_{0.75}$As barrier (100nm thick), which was MBE-grown on an n+ GaAs substrate. The counter-doped barrier results in band bending, which provides additional confinement of the holes to the interface region (Fig. 3 b), and results in a reduction in inelastic tunneling. The superconductor-normal (S-N) junction is in the direction of growth in the semiconductor, and along the *c*-axis of Bi-2212. However after the S-N vertical junction, the transport inside GaAs is between the contacts separated in the lateral direction on top of the GaAs sample (Fig. 1 inset, Fig. 3 a). In the QW-based junction with additional carrier confinement obtained with the counter-doping based band bending, the transport is even more well-defined to be in plane (Fig. 3 b).

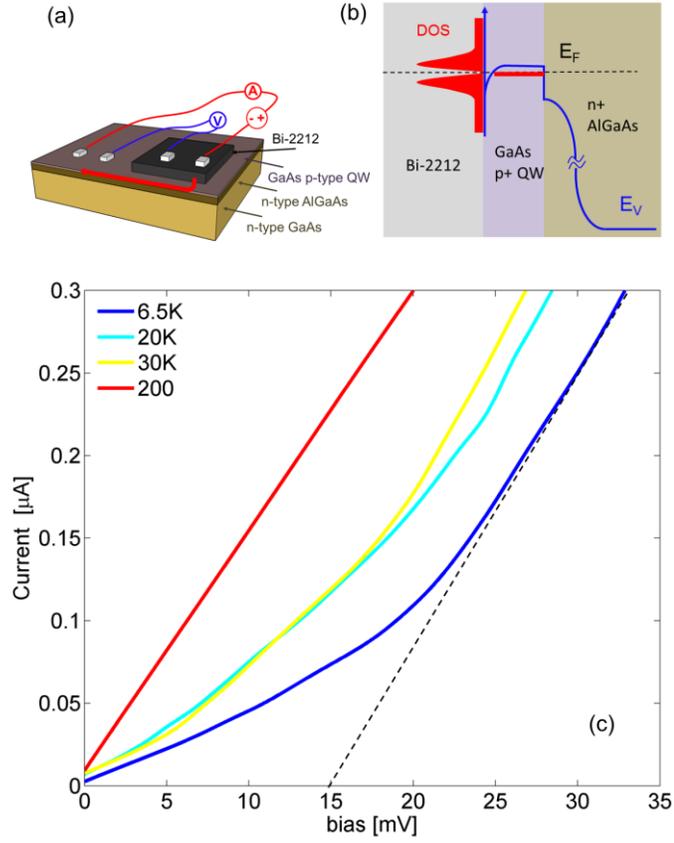

**Figure 3.** (a) Schematic drawing of the Bi-2212/GaAs/AlGaAs device, (b) energy diagram of the device, (c) Measured DC I-V characteristics for the Bi-2212/GaAs/AlGaAs junction at various temperatures. The dashed line indicates a linear I-V dependence coinciding with the 6.5K measurement at higher bias to indicate the excess voltage.

Below $T_c$, the measured I-V characteristics for a Bi-2212/GaAs/AlGaAs QW junction exhibit the nonlinear I-V dependence and excess voltage induced by quasiparticle tunneling (Fig. 3 c). Above $T_c$, the I-V dependence is linear, as expected from a superconducting tunnel diode. However, the nonlinearity below $T_c$ in a QW-based junction is much stronger than in the bulk GaAs junction (Fig. 1), and the excess voltages $V_e$ are larger than 10 mV. The enhanced excess voltages in QW junctions are caused by the reduced scattering rate in the two-dimensional QW system, which results in a stronger effect of the superconducting gap of Bi-2212 on the tunneling spectra.

To model the effect quantitatively, we calculated the *c*-axis Bi-2212/GaAs and Bi-2212/GaAs/AlGaAs QW tunneling spectra using an extension of the Blonder-Tinkham-Klapwijk superconductor-normal interface formalism [44], developed for anisotropic superconductors [29]. The ratio of the differential conductance $[dI/dV]_S$, to the normal state conductance $[dI/dV]_N$ is given by the half-sphere integration over solid angle $\Omega$:

$$\sigma(E) = \frac{\int d\Omega \sigma_N \cos\theta_N \sigma_R(E)}{\int d\Omega \sigma_N \cos\theta_N} \quad (1)$$

where $E$ and $\theta_N$ are the quasiparticle energy and incidence angle in the normal material respectively, $\sigma_N$ is the conductance from normal to normal material with the same geometry, and

$$\sigma_R(E) = \frac{1 + \sigma_N |\kappa_+|^2 + (\sigma_N - 1)|\kappa_- \kappa_+|^2}{\left|1 + (\sigma_N - 1)|\kappa_- \kappa_+|\exp(i\varphi_- - i\varphi_+)\right|^2} \quad (2)$$

with

$$\kappa_\pm = \left[ E + i\Gamma - \sqrt{(E+i\Gamma)^2 - |\Delta_\pm|^2} \right] \Big/ |\Delta_\pm| \quad (3)$$

and $\Delta_\pm = |\Delta_\pm|\exp(i\varphi_\pm)$ are electron-like and hole-like quasiparticle effective pair potentials with the corresponding phases $i\varphi_\pm$. The calculations were performed with equal maximal values of the gap and the barrier for both the bulk and the QW based junctions, while the scattering-induced energy broadening $i\Gamma$ is included in the calculation as a fitting parameter. The origin of this broadening term in tunneling experiments is not always clear. However, recent studies have shown that when a superconductor is coupled to a dissipative environment, the Dynes broadening can be explained mostly by quasiparticles exchanging energy with the environment during tunneling, and a recent experiment has demonstrated the effect of electromagnetic radiation on the Dynes broadening [41].

In our case, the QW and the bulk GaAs are the primary source of inelastic scattering in the tunneling, acting therefore as the environment. The quasiparticles can exchange energy with phonons or charge carriers in the semiconductor during tunneling. Scattering in the semiconductor, therefore, can contribute a significant portion of the broadening, $\Gamma$. Other dissipative processes during tunneling could also contribute to the broadening, however the main difference between the QW and the bulk GaAs experiments is the dimensionality of the semiconductor affecting the dissipative scattering rates.

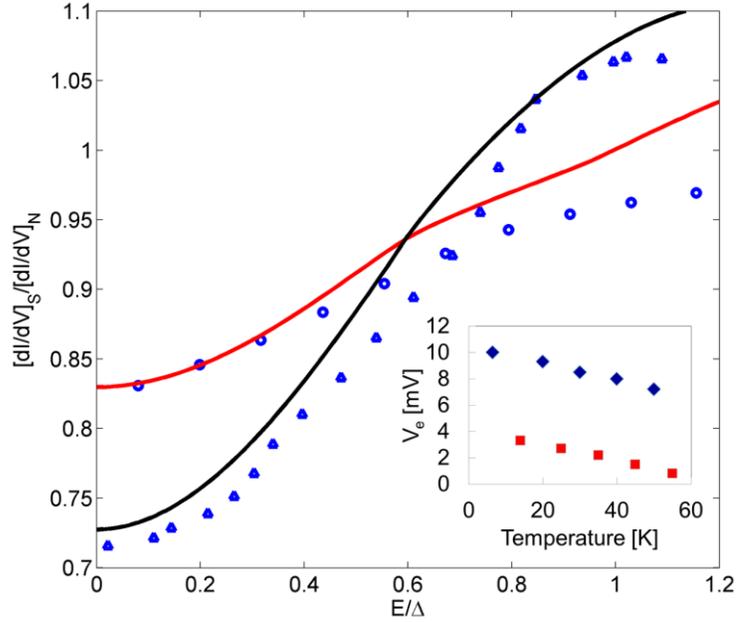

**Figure 4.** AC differential conductance $[dI/dV]_S$ at 14K, divided by the normal state conductance $[dI/dV]_N$ for the Bi-2212/GaAs junction measured (circles) and calculated (solid red) with the $\Gamma$=15meV, and for Bi-2212/GaAs/AlGaAs junction measured (triangles) and calculated (solid black) with the $\Gamma$ = 6meV. The inset is the excess voltage vs. temperature in a bulk GaAs device (red squares) and in the QW-based device (blue diamonds).

Our calculations according to this model are consistent (within a few percent) with the experimental results (Fig. 4). The difference between the density of states in the two different kinds of normal materials, namely bulk GaAs and the QW, is accounted for by normalizing each to the normal conductance above $T_c$. The large quasiparticle energy broadening in bulk GaAs significantly reduces the depth of the superconducting gap, resulting in a smaller excess voltage (Fig. 1). In contrast, the reduced scattering rate in the two-dimensional GaAs/AlGaAs quantum well system enhances both the dip in the differential conductance (Fig. 4) and the excess voltage (Fig. 3 c). The larger nonlinearity in the QW-based device relative to the bulk device persists at much higher temperatures resulting in larger excess voltages (Fig. 4 inset).

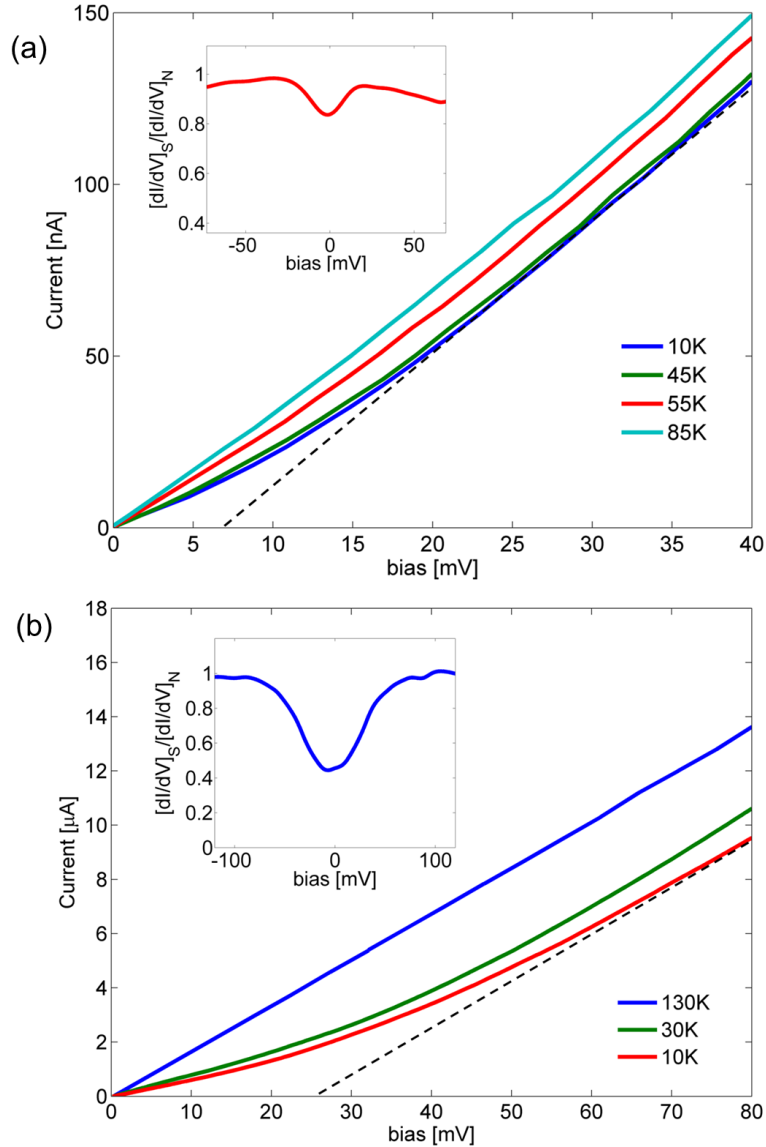

**Figure 5. (a)** Measured DC I-V characteristics for the Bi-2212/graphite junction at various temperatures. The dashed line indicates a linear I-V dependence coinciding with the 10K measurement at higher bias to indicate the excess voltage. The inset is the measured differential conductance $[dI/dV]_S$ at 10K, divided by the normal state conductance $[dI/dV]_N$ for the Bi-2212/graphite junction. **(b)** Measured DC I-V characteristics for the Bi-2212/ Bi$_2$Te$_3$ junction at various temperatures. The dashed line indicates a linear I-V dependence coinciding with the 10K measurement at higher bias to indicate the excess voltage. The inset is the measured differential conductance $[dI/dV]_S$ at 10K, divided by the normal state conductance $[dI/dV]_N$ for the Bi-2212/ Bi$_2$Te$_3$ junction.

To demonstrate the broad applicability of our technology, we have constructed superconducting tunnel junctions from additional materials. The third type of material we used to construct a superconducting-normal junction was bulk graphite. A mechanically-bonded junction of Bi-2212 on graphite exhibits excess voltage (Fig. 5a) similar to that obtained in GaAs based junctions. The Bi-2212 in this junction is also slightly underdoped resulting in correspondingly smaller gap (Fig. 5a inset) and slightly lower T$_c$. The forth material we employed for the normal side of the high-T$_c$ diode was Bi$_2$Te$_3$. In this experiment, an optimally doped Bi-2212 was mechanically-bonded with Bi$_2$Te$_3$ [45]. The larger gap of an optimally doped Bi-2212 (Fig. 5 b inset) results in enhanced nonlinearity and excess voltage as large as 25mV (Fig. 5 b).

The mechanical bonding method presented here was reproduced in several different hetero-structures. These results suggest wide applicability of the method; however they do not yet constitute proof of universality. Extensions of our approach to other material classes have to be carefully studied, and will have to address the need to achieve controllable and reproducible interfaces in each specific case while employing the bonding technique.

In conclusion, we have demonstrated hybrid high-T$_c$ superconductor-semiconductor devices, which were fabricated by our newly-developed method of mechanically-bonded planar junctions. This technique is substantially different from previous studies of low-T$_c$ semiconductor-superconductor hybrid devices, providing a unique solution for combinations of various novel materials unobtainable by conventional means. This method allows simple room-temperature fabrication of devices based on various easily cleavable crystals including combinations of high-T$_c$ superconductor such as Bi-2212 with graphite, Bi$_2$Se$_3$ and Bi$_2$Te$_3$.

The devices demonstrated here are junctions constructed from Bi-2212 combined with bulk GaAs, GaAs/AlGaAs quantum wells, graphite or Bi$_2$Te$_3$ acting as superconducting tunnel diodes. Tunneling spectra characterization of the hybrid junctions exhibits excess voltage and nonlinearity in DC I-V measurements as well as decrease in conductance due to the Bi-2212 superconducting gap in AC differential conductance measurements, in good agreement with theory. Additional junctions constructed from Bi-2212 combined with graphite or $Bi_2Te_3$ also show excess voltage and nonlinearity. These results enable practical high-$T_c$ hybrid superconductor-semiconductor devices with new insights into fundamental studies of novel materials as well as applications in optoelectronics and quantum technologies.

The work at the University of Toronto was supported by the Natural Sciences and Engineering Research Council of Canada, the Canadian Foundation for Innovation, and the Ontario Ministry for Innovation. The crystal growth at Princeton was supported by the US National Science Foundation, grant number DMR-0819860.